\documentclass[ reprint,nofootinbib,superscriptaddress,amsmath,amssymb,aps]{revtex4-1}
\usepackage{graphicx}
\usepackage{rotating}
\usepackage{dcolumn}
\usepackage{bm}
\usepackage{color}
\usepackage{mathptmx, textcomp}
\usepackage[latin1]{inputenc}
\usepackage{braket}
\usepackage[colorlinks,linkcolor=magenta,anchorcolor=cyan,citecolor=blue]{hyperref}
\usepackage[multiple]{footmisc}
\newcommand{\fig}[1]{Fig.\ref{#1}}
\def\be{\begin{equation}}
\def\ee{\end{equation}}
\def\ba{\begin{eqnarray}}
\def\ea{\end{eqnarray}}

\def\lf{\left}
\def\rt{\right}

\newcommand{\eq}[1]{(\ref{#1})}

\def\n{\nonumber}\def\lf{\left}\def\rt{\right}\def\q{\theta}   \def\y {\psi}   \def\p {\pi} \def\a {\alpha}  \def\d {\delta} \def\f {\phi} \def\g {\gamma}   \def\k {\kappa}   \def\x {\xi}  \def\b {\beta}   \def\p {\partial} \def \inf {\infty}  
 \def\W{\Omega}     \def\S {\Sigma}  \def\F {\Phi}  \def\L {\Lambda}    \def\grad{\nabla}\def\.{\cdot}

\begin{document}

\title{General Smarr relation and first law of Nutty dyonic black hole}
\author{Zhaohui Chen}
\email{chenzhaohui@mail.bnu.edu.cn}
\affiliation{Department of Physics, Beijing Normal University, Beijing, 100875, China}
\author{Jie Jiang}
\email{Corresponding author. jiejiang@mail.bnu.edu.cn}
\affiliation{Department of Physics, Beijing Normal University, Beijing, 100875, China}

\date{\today}

\begin{abstract}
In this paper, we investigate the thermodynamics of the charged NUT black hole with the Misner strings present in the Einstein-Maxwell gravity. We show that the Misner charge $N$ can be obtained by performing Komar integration over the Misner strings in the presence of electric and magnetic charges and the corresponding Smarr relation and first law can also be naturally established based on this geometric perspective. Besides, we demonstrate that the electric and magnetic charges appearing in the first law and Smarr relation are contingent on the special choice of gauge freedom for the electric and magnetic potentials. We re-derive two different versions of thermodynamics according to distinct choices of gauge and furthermore, we try to formulate more general thermodynamic laws under other arbitrary choices of gauge and deduce the conditions necessary to satisfy the Smarr relations and first laws simultaneously.
\end{abstract}
\maketitle
\section{Introduction}
As one of the most interesting solutions of general relativity, the Taub-NUT spacetime \cite{Taub, Newman} has engendered lots of perplexing problems since the birth. In this spacetime exist two kinds of Killing vectors which respectively correspond to the Schwarzschild-like horizon and Misner string singularities on the north and south pole axes due to the existence of NUT parameter $n$. The Misner strings have also resulted in appearance of spacetime regions with closed timelike curves in its vicinity.

To eschew these issues, Misner proposed that imposing the periodicity of time coordinate makes the string singularities unobservable \cite{CWM}. Since then, the following studies on the Taub-NUT spacetime mostly focused on Euclidean case. However, this manipulation also brings about many new problems. For instance, it leads to the existence of closed timelike curves everywhere and makes the spacetime geodesically incomplete \cite{CWM,Hawking,Hajicek} and moreover, when establishing thermodynamics of this spacetime, the NUT parameter $n$ is no longer a valid independent variable  but rather a function of Schwarzschild-like horizon radius $r_h$ due to the identification between the standard temperature formula and the reciprocal of the periodicity of Euclidean time coordinate \cite{Stelea}. This also results in the possibility that the thermodynamic volume is negative, and the entropy obtained is not in accord with the Bekenstein-Hawking area law \cite{Johnson1,Johnson2}.

 However, recent studies have demonstrated that the removal of Misner strings by imposing the periodicity of time coordinate in Euclidean setup is unnecessary in fact. Cl\'{e}ment et al. \cite{Clement1,Clement2,Clement3} illustrated that the Misner string singularities are far less problematic than previously thought and they argued that Lorentzian Taub-NUT solution without imposing the Misner time periodicity condition is geodesically complete. At the same time, despite the existence of regions with closed timelike curves, causality is not violated at all for geodesic observers, implying that Lorentzian Taub-NUT spacetimes with the Misner strings present may be physical in essence.

Recently, D. Kubiz\v{n}\'{a}k et al. have formulated a reasonable thermodynamics of Lorentzian Taub-NUT-AdS spacetime with the Misner strings present \cite{DK1}. They treated the NUT parameter $n$ as independent variable, and thus introduced a new pair of conjugate thermodynamic quantities $\psi$--$N$, where $\psi$ is called the Misner potential and its conjugate term $N$ is named as the Misner gravitational charge. Such a new first law and Smarr relation just take the following forms:
\begin{align}
\delta M&=T\,\delta S+\psi\,\delta N+V\,\delta P, \n\\
M&=2(TS-VP+\psi N).
\end{align}
Here, $M$ represents the mass of this spacetime, $T$ is the temperature of  Schwarzschild-like horizon, $S$ is the Bekenstein-Hawking entropy, $P$ is the cosmological constant pressure and $V$ is the conjugate quantity of it. In this case, the cosmological constant is treated as a state parameter of the thermodynamic system. And this can be understood since the cosmological constant can be induced from a totally antisymmetric tensor field sourced by a membrane \cite{Brown1, Brown2}. Originally, they just obtained the free energy by calculating Euclidean action \cite{Emparan} to identify the several thermodynamic terms and further determined the Misner charge $N$. Later, geometrical interpretation on the two novel quantities $\psi$ and $N$ was assigned \cite{DK3} and it was shown that the Misner potential is exactly the surface gravity of the Misner strings and the Misner charge just corresponds to the Komar integration over the tubes surrounding the Misner strings.

However, when extended to the charged Taub-NUT-AdS spacetime in the Einstein-Maxwell gravity \cite{Brill,Demianski,NA}, in this case the electric and magnetic charges which were called ``Nutty dyon" both exist and the magnitudes of them depend on the radius of the 2-dimensional sphere (see (\ref{qeqm}) below). Thus, an important problem occurs in formulation of thermodynamics, namely how to choose the proper electric charge $q_e$ and magnetic charge $q_m$ appearing in the first law and Smarr relation as thermodynamic quantities. Ref. \cite{DK2} gave two different types of thermodynamics of Nutty dyonic black hole, and one is that the asymptotic electric charge and horizon magnetic charge are regarded as independent thermodynamic quantities, while as the other case, the horizon electric charge and the asymptotic magnetic charge are considered. Hence, two types of corresponding Smarr relations can be written as follows:
\begin{align}
M&=2(TS-VP+\y N^{(1)})+\f_e Q_e+\f_m Q_m^{(+)}\,,\label{smarr1}\\
M&=2(TS-VP+\y N^{(2)})+\f_e Q_e^{(+)}+\f_m Q_m\label{smarr2}\,,
\end{align}
Conspicuously, Misner charges for the charged case differ in these two distinct versions of thermodynamics. However, the expressions of Misner charges are still determined by means of calculating Euclidean action for free energy \cite{DK2} and the similar Komar integrations of them for Nutty dyonic black holes are not given in the previous literature. In addition, it deserves our deep exploration that why one can formulate two different versions of thermodynamics in this spacetime and whether we can give an interpretation from geometrical notions.

So our purpose in this paper is to construct thermodynamic laws of this Nutty dyonic black hole from the Komar integral perspective and recover the corresponding Misner charges given in \cite{DK2}. Meanwhile, we illustrate that formulations of two different versions of thermodynamics are actually relevant to the gauge choices of electric and magnetic potentials and moreover, we attempt to obtain more general types of thermodynamics based on arbitrary gauge choices from this perspective.

Our paper is organized as follows. In the next section, we briefly review the rudiments about the Taub-NUT-AdS spacetime and enumerate some basic thermodynamic quantities for Nutty dyonic black hole. In Sec. \ref{3} we present the geometrical philosophy to derive the Smarr relations and discuss how different gauge choices of electric and magnetic potentials influence the formulations of thermodynamics. Sec. \ref{4} is devoted to calculations of the Misner charges in two special distinct gauges. In Sec. \ref{5}, we derive more general Smarr relations for Nutty dyonic black hole under arbitrary choices of gauge and give essential conditions to satisfy the first law. Finally, conclusions are presented in Sec. \ref{6}.

\section{Geometry of charged TAUB-NUT-AdS spacetime}
In this paper, we consider the charged Lorentzian Taub-NUT-AdS solution of the $4$-dimensional Einstein-Maxwell gravity, where the action can be written as
\ba
I=\frac{1}{16\pi}\int_{M}d^4x\sqrt{-g}\lf(R-2\L-F_{ab}F^{ab}\rt)\,,
\ea
in which $\bm{F}=d\bm{A}$ is the electromagnetic strength and $R$ is the Ricci scalar of the spacetime. The equations of motion are given by
\ba\label{eom}\begin{aligned}
R_{ab}-\frac12Rg_{ab}+\L g_{ab}&=T_{ab}\,,\\
d\bm{G}=0,
\end{aligned}\ea
with the energy-momentum tensor of the Maxwell field
\begin{align}
T_{ab}=F_{ac}{F_b}^c+G_{ac}{G_b}^c \label{Tab}\,,
\end{align}
and
\begin{align}
{\bm G}=\star{\bm F}\,.
\end{align}
The electric and magnetic charges inside a $2$-dimensional surface $S^2$ can be defined as
\ba\begin{aligned}\label{QeQm}
q_e[S^2]=\frac{1}{4\pi}\int_{S^2}\bm{G}\,,\ \ \ q_m[S^2]=\frac{1}{4\pi}\int_{S^2}\bm{F}\,.
\end{aligned}\ea
According to the equations of motion \eq{eom}, the Taub-NUT-AdS solution can be read off \cite{Johnson2,NA}
\ba\begin{aligned}\label{A}
ds^2=&-f(r)\left(dt+2n\cos\theta\,d\phi\right)^2+\frac{dr^2}{f(r)} \\
&+\left(r^2+n^2\right)\left(d\theta^2+\sin^2\theta\,d\phi^2\right)\\
{\bm A}=&-\lf[h(r)-h_0\rt]dt+2 h(r) n\cos\theta\,d\phi
\end{aligned}\ea
where
\begin{align}
f(r)=\frac{r^2-2m r-n^2+4n^2 g^2+e^2}{r^2+n^2}-\frac{3n^4-6n^2r^2-r^4}{l^2(r^2+n^2)} \n
\end{align}
is the blackening factor,
\begin{align}
h(r)=\frac{e r}{r^2+n^2}+\frac{g(r^2-n^2)}{r^2+n^2}\,,
\end{align}
and $h_0$ is some arbitrary constant which reflects the gauge freedom of the electromagnetic field. Besides, $n$ and $m$ are the NUT and mass parameters, $e$ and $g$ are the electric and magnetic parameters, and $l$ is the AdS radius with $\Lambda=-3/l^2$.

Performing the integration over a sphere of radius $r$ according to Eq. \eq{QeQm} , we can obtain the electric and magnetic charges inside this sphere
\ba\begin{aligned}
q_e(r)&=\frac{e(r^2-n^2)-4g r n^2}{r^2+n^2}\,,\\ q_m(r)&=\frac{2n[er+g(r^2-n^2)]}{r^2+n^2}\,. \label{qeqm}
\end{aligned}\ea
One can note that disparate with common results of the dyonic solutions, the charges vary with the radius of the sphere $S^2$, which can be easily understood since the sphere $S^2$ is not a close surface due to the existence of the Misner singularities. And two kinds of special values of these charges are respectively the asymptotic $(r\to\inf)$ charges:
\ba
Q_e=q_e(r\to\inf)=e\,,\ \ \ Q_m=q_m(r\to\inf)=2gn\,,
\ea
and the horizon $(r=r_h)$ charges:
\ba
Q_e^{(+)}=q_e(r_h)\,,\ \ \ Q_m^{(+)}=q_m(r_h)\,,
\ea
where $r_h$ is the largest root of $f(r_h)=0$. By using the conformal method \cite{AS}, the mass of this black hole is obtained as
\ba
M=m\,,
\ea
while the total angular momentum of the spacetime vanishes. According to the line element \eq{A}, the Killing vector of the horizon is manifestly given by
\begin{align}
k^a=\left(\frac{\p}{\p t}\right)^a\,.
\end{align}
And the temperature of this Schwarzschild-like Killing horizon is given by the surface gravity
\ba\begin{aligned}
T&=\frac{\k_h}{2\pi}=\frac{f'(r_h)}{4\pi}\\
&=\frac{1}{4\pi r_h}\left(1+\frac{3(r_h^2+n^2)}{l^2}-\frac{e^2+4n^2 g^2}{r_h^2+n^2}\right)\,.
\end{aligned}\ea
The entropy can be identified  with the horizon area law:
\begin{align}
S=\frac{A}{4}=\pi\left(r_h^2+n^2\right)\,.
\end{align}
Moreover, there are yet other Killing horizons present in the spacetime, which are separately located on the north and south pole axes with the Killing vectors
\begin{align}
\xi^a_{\pm} =k^a\mp\frac{1}{2 n}\varphi^a
\end{align}
with $\varphi^a=\left(\p/\p\phi\right)^a$. The corresponding surface gravity can be calculated by the standard formula
\begin{align}
\kappa^2_\xi=\frac12\grad_a\xi_b\grad^a\xi^b\,,
\end{align}
which gives the Misner potentials
\begin{align}
\psi=\psi_\pm=\frac{\kappa_\x}{2\pi}=\frac{1}{8\pi n}\,. \label{psi}
\end{align}
In what follows we will identify the conjugate pairs appearing in the Smarr relations with the Komar integrations and give different types of electric and magnetic terms from the geometric perspective.

\section{Smarr relation with electric and magnetic charges}\label{3}
In \cite{DK2}, the authors gave the consistent first law and Smarr relation of the Nutty dyonic black hole containing the electric and magnetic charges based on the thermodynamic consideration. Inspired by the uncharged case \cite{DK3} and Taub-NUT solution with rotation \cite{DK4}, in this section we would like to derive a general Smarr relation and then apply it to reexpress the thermodynamic quantities obtained in \cite{DK2} from the Komar integral perspective. The key point to achieve it is the decomposition of the mass \cite{Kastor}. Now, for the Killing vector $k^a$, we have
\begin{align}
d(\star{\bm k})=0\,,
\end{align}
which implies that there locally exists a $2$-form ${\bm\omega}$, such that
\begin{align}
\star{\bm k}=d{\bm\omega}\,.\label{domega}
\end{align}
Then, together with the equation of motion \eq{eom}, utilizing the facts that $k^a$ is a Killing vector and the sum of cyclic permutations of the last three indices in ${R^a}_{bcd}$ vanishes, we thus have the following identity:
\begin{align}
\grad_a \grad^a k^b&=-{R^b}_ak^a \n\\
&=-\Lambda k^b-T^{ab}k_a\,. \label{grad}
\end{align}

\begin{figure}
\begin{center}
\includegraphics[width=0.28\textwidth,height=0.21\textheight]{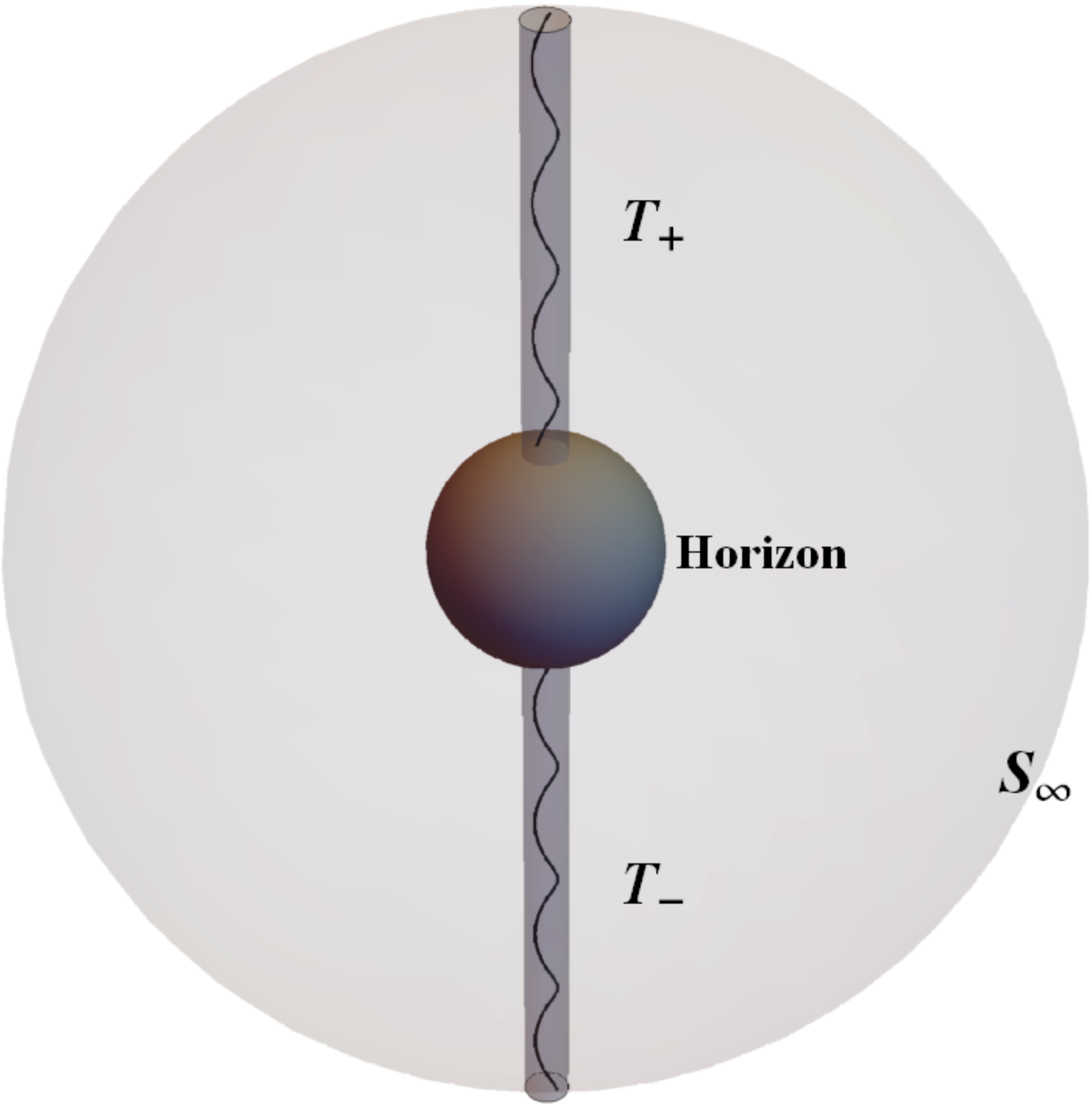}
\caption{{\bf Taub--NUT--AdS boundaries: Misner tubes \cite{DK3}}, as a result of the effect of NUT parameter $n$, besides the standard boundaries on the Schwarzschild-like horizon $S_h$ and at infinity $S_\infty$, the two Misner tubes $T_\pm$ of radius $\epsilon$ surrounding the symmetric strings are located at the north and south pole axes, $\cos\theta=\pm1$.}\label{fig}
\end{center}
\end{figure}

We observe that for the charged case, energy-momentum term will be devoted to constructing the the electric and magnetic charge terms in the Smarr relation. Thus, we need to transform this term into differential forms using the expression \eq{Tab} of $T^{ab}$ and then give the Komar integrals of electric and magnetic terms. Moreover, from the equation of motion $d\bm{G}=0$ of the electromagnetic field, we can see that the on-shell value of $\bm{G}$ is a closed $2$-form, which implies that there exists a $1$-form ${\bm B}$ such that ${\bm G}=d{\bm B}$. We can note that the new-defined vector potential $\bm{B}$ also allows a gauge freedom. So, substituting  Eq. \eq{Tab} into the above expression, the second term becomes
\begin{align}
&T^{ab}k_a=k_a{F^a}_cF^{bc}+k_a{G^a}_cG^{bc} \n\\
&=F^{bc}k_a\grad^a A_c-F^{bc}k_a\grad_c A^a+G^{bc}k_a\grad^a B_c-G^{bc}k_a\grad_c B^a \n\\
&=-F^{bc}A_a\grad_c k^a-F^{bc}k^a\grad_c A_a-G^{bc}B_a\grad_c k^a-G^{bc}k^a\grad_c B_a \n\\
&=-F^{bc}\grad_c \Phi_e-G^{bc}\grad_c \Phi_m \n\\
&=\grad_a\left(F^{ab}\Phi_e+G^{ab}\Phi_m\right)\,,\label{Tk}
\end{align}
where we have used the fact that Lie derivatives of vector potentials ${\bm A}$ and ${\bm B}$ along the vector $k^a$ vanish, namely ${\mathcal L}_{\bm k}\bm{A}=0$ and ${\mathcal L}_{\bm k}\bm{B}=0$, and denote the contractions of Killing vector $k^a$ with two vector potentials as
\begin{align}
\Phi_e=k^aA_a\,, \quad \Phi_m=k^aB_a\,, \label{kaba}
\end{align}
which can be regarded as the electric and magnetic potentials of Maxwell field. And the counterparts of the black hole can be defined as
\ba
\f_e=\F_e(\inf)-\F_e(r_h), \quad \f_m=\F_m(\inf)-\F_m(r_h).\label{phiephim}
\ea
Then, Eq. (\ref{grad}) together with (\ref{Tk}) yields
\begin{align}
\grad_a\left(\grad^a k^b+F^{ab}\Phi_e+G^{ab}\Phi_m\right)=-\L k^b\,.
\end{align}
Using the language of differential forms, we have
\begin{align}
d\left(\star d{\bm k} +2\Phi_e{\bm G}-2\Phi_m{\bm F}\right)+2\L\star{\bm k}=0\,,
\end{align}
which implies
\begin{align}
d\left(\star d{\bm k} +2\Phi_e{\bm G}-2\Phi_m{\bm F}+2\L{\bm\omega}\right)=0\,,\label{weifen}
\end{align}
Integrating over a $3$-dimensional hypersurface $\Sigma$ and using the Stokes theorem, it can be expressed by some boundary quantities,
\begin{align}
0&=\int_{\Sigma}d\left(\star d{\bm k} +2\Phi_e{\bm G}-2\Phi_m{\bm F}+2\L{\bm\omega}\right) \n\\
&=\int_{\p\Sigma}\left(\star d{\bm k} +2\Phi_e{\bm G}-2\Phi_m{\bm F}+2\L{\bm\omega}\right)\,.\label{int1}
\end{align}
For the normal black hole, the hypersurface $\S$ is bounded by the spheres $S_h$ on the horizon and $S_\inf$ at asymptotic infinity. However, when the Misner strings are present, there exist the Misner string singularities which locate at $\q=0$ and $\q=\pi$. Then, the decomposition will depend on the Misner strings. Therefore, we also need introduce two Misner string tubes $T_+$ and $T_-$ which are located at $\q=\epsilon$ and $\q=\pi-\epsilon$ (see \fig{fig}) with infinitesimal parameter $\epsilon$, respectively. After taking into account the orientation of the boundaries, we have
\begin{align}
\p\Sigma=T_++S_\inf-T_--S_h\,.
\end{align}

Thus, the integral \eq{int1} can be decomposed into the following integrations over several pieces of boundaries:
\begin{align}
0=&\int_{S_\infty}\left(\star d{\bm k}+2\L{\bm\omega}\right)-\int_{S_h}\star d{\bm k} \n\\
&+2\int_{S_\infty} (\Phi_e{\bm G}-\Phi_m{\bm F})-2\int_{S_h}(\Phi_e{\bm G}-\Phi_m{\bm F}) \n\\
&-2\L\left(\int_{S_h}{\bm\omega}+\Omega_+^{(r=r_h)}-\Omega_-^{(r=r_h)}\right) \n\\
&+\int_{T_+}\left(\star d{\bm k}+2\Lambda\Omega_+^{(r=\infty)}+2\Phi_e{\bm G}-2\Phi_m{\bm F}\right) \n\\
&-\int_{T_-}\left(\star d{\bm k}+2\Lambda\Omega_+^{(r=\infty)}+2\Phi_e{\bm G}-2\Phi_m{\bm F}\right)\,.\label{integral}
\end{align}
In the same way as the uncharged case \cite{DK3}, we also denote
\begin{align}
\int_{T_\pm}{\bm\omega}=\Omega^{(r=\infty)}_\pm-\Omega^{(r=r_h)}_\pm\,, \label{omegaintegral}
\end{align}
splitting the integral into the $r\to \inf$ and $r=r_h$ parts. From the Eq. (\ref{integral}) we can see that the existence of electromagnetic field provides a modification for the Misner charge apart from furnishing the electric and magnetic terms.  By identifying the thermodynamic mass $M$, cosmological constant pressure $P$ and the product of entropy and temperature with corresponding Komar integrals \cite{Kastor},
\ba\begin{aligned}\label{MPTS}
M&=-\frac{1}{8\pi}\int_{S_\infty}\left(\star d{\bm k}+2\Lambda\omega\right), \quad P=-\frac{\L}{8\pi}\,,\\
TS&=-\frac{1}{16\pi}\int_H\star d{\bm k}\,,
\end{aligned}\ea
Eq. \eq{integral} can be simplified as
\begin{align}
M=&2\lf(TS-PV+\y N^\F\rt)+\lf[\F_e q_e-\F_m q_m\rt]^\inf_{r_h} \,.\label{dec}
\end{align}
With the similar consideration as \cite{DK3}, we also define the thermodynamic volume and Misner gravitational charge as
\ba\begin{aligned}
V&=-\left(\int_{S_h}{\bm\omega}+\Omega_+^{(r=r_h)}-\Omega_-^{(r=r_h)}\right)\,,\\
N^\F&=N_+^\F+N_-^\F
\end{aligned}\ea
with
\ba\begin{aligned}\label{NFF}
N_\pm^\F&=\pm\frac{1}{16\pi \y}\int_{T_\pm}\left(\star d{\bm k}+2\L\Omega_{\pm}^{(r=\infty)}+2\Phi_e{\bm G}-2\Phi_m{\bm F}\right). \label{gmisner}
\end{aligned}\ea

So far, we have derived a more general Smarr relation for Nutty dyonic black hole. Normally, for some black holes with charges $q_e$ and $q_m$ being constants, we know that the gauge freedoms of $\Phi_e$ and $\Phi_m$ are trivial from Eq. \eq{phiephim}. For this case since $q_e$ and $q_m$ are no longer constants but rather functions of the radius $r$ of sphere $S^2$ as demonstrated in Eq. (\ref{qeqm}), so different gauge choices of the electric and magnetic potentials $\F_e$ and $\F_m$ will give distinct electric and magnetic terms appearing in the Smarr relations and as is shown in Eq. \eq{gmisner}, corresponding Misner charges $N^{\Phi}$ will also make variations accordingly. Thus, we realize that in order to obtain the similar decompositions as the Smarr relations \eq{smarr1} and \eq{smarr2}, we need choose some special gauges. If choosing the gauge $\F_e=\F_e^{(1)}, \F_m=\F_m^{(1)}$ such that
\ba\label{gauge1}
\F_e^{(1)}(r_h)=0\,,\ \ \F_m^{(1)}(\inf)=0\,,
\ea
we have $ \F_e^{(1)}(\inf)=\f_e$, $\F_m^{(1)}(r_h)=-\f_m$, and this choice ensures that asymptotic electric charge $Q_e$ and horizon magnetic charge $Q_m^{(+)}$ appear in the first law and Smarr relation. Then Eq. \eq{dec} becomes
\begin{align}\label{MT1}
M=&2\lf(TS-PV+\y N^{(1)}\rt)+\f_e Q_e+\f_m Q_m^{(+)} \,,
\end{align}
where $N^{(1)}=N^{\F^{(1)}}$ is the value of Misner charge under the above gauge \eq{gauge1}. Equation. \eq{MT1} gives the same decomposition as the Smarr relation \eq{smarr1} by the Komar integral.

Based on  another choice of gauge, we can formulate an alternative Smarr relation \eq{smarr2}, where the electric charge is taken on the horizon and the magnetic counterpart at infinity. Here on the contrary to Eq. (\ref{gauge1}), we should choose the gauge $\F_e=\F_e^{(2)}, \F_m=\F_m^{(2)}$ such that
\ba\label{gauge2}
\F_m^{(2)}(r_h)=0\,,\ \ \F_e^{(2)}(\inf)=0\,.
\ea
Thus, we have $\Phi_e^{(2)}(r_h)=-\phi_e$, $\Phi_m^{(2)}(\infty)=\phi_m$ and the decomposition becomes
\begin{align}\label{MT2}
M=&2\left(TS-PV+\y N^{(2)}\right)+\f_e Q_e^{(+)}+\f_m Q_m \,,
\end{align}
which is exactly the Smarr relation (\ref{smarr2}). And here, $N^{(2)}$ is the value of $N^\Phi$ under the gauge \eq{gauge2}. The above discussion shows that the corresponding Misner charges obtained in \cite{DK2} can be expressed as some Komar integrations \eq{NFF} in two special choices of gauge \eq{gauge1} and \eq{gauge2}.

Finally, in order to ensure the rationality of these quantities, we should check the validity of the first law of thermodynamics, i.e.,
\ba\label{firstlaw}\begin{aligned}
\d M=&T\d S+V\d P+\y\d N^{(1)}+\f_e \d Q_e+\f_m \d Q_m^{(+)} \,,\\
\d M=&T\d S+V\d P+\y\d N^{(2)}+\f_e \d Q_e^{(+)}+\f_m \d Q_m \,.
\end{aligned}\ea
\section{First law with variable Misner charges under some special gauge choices}\label{4}
In the former section, we elaborate on the basic strategy for establishing the thermodynamics of Nutty dyonic black hole from geometric perspective and show that different gauge choices of $\Phi_e$ and $\Phi_m$ will give distinct types of Smarr relation. From Eq. \eq{gmisner}, we know that Misner charges are also subject to influences from the gauge freedoms as well as the electric and magnetic terms. And below, we are devoted to calculating the Komar integrations of Misner charges and other thermodynamic quantities to recover the results in Ref. \cite{DK2} in two special gauges. Besides, the corresponding first law \eq{firstlaw} will be examined. According to Eq. \eq{A}, we immediately obtain electromagnetic field tensor
\begin{align}
{\bm F}&=-\frac{e(n^2-r^2)+4n^2gr}{(r^2+n^2)^2}\,dr\wedge dt+\frac{2n[er+g(r^2-n^2)]}{r^2+n^2} \n\\
\times&\sin\theta d\theta\wedge d\phi-2n\frac{e(n^2-r^2)+4n^2gr}{(r^2+n^2)^2}\cos\theta\,dr\wedge d\phi. \n
\end{align}
Then, we can instantly obtain
\begin{align}
{\bm G}&=\frac{2n[g(n^2-r^2)-er]}{(r^2+n^2)^2}\,dr\wedge dt+\frac{e(r^2-n^2)-4n^2gr}{r^2+n^2} \n\\
\times&\sin\theta d\theta\wedge d\phi+\frac{4n^2[g(n^2-r^2)-er]}{(r^2+n^2)^2}\cos\theta\,dr\wedge d\phi. \n
\end{align}
From this, it is not difficult to verify the new-defined vector potential
\begin{align}
{\bm B}=[p(r)-p_0]dt+2np(r)\cos\theta\,d\phi
\end{align}
with
\begin{align}
p(r)=\frac{4n^2gr-e(r^2-n^2)}{2n(r^2+n^2)}.
\end{align}
and arbitrary constant $p_0$ which should be determined when we fix the gauge. At the same time, we evaluate the Hodge dual of $k^a$, namely
\begin{align}
\star d{\bm k}=&-\frac{2fn}{r^2+n^2}\,dr\wedge(dt+2n\cos\theta\,d\phi)\n\\
&-\sin\theta(r^2+n^2)f'\,d\theta\wedge d\phi.
\end{align}
Then from Eq. (\ref{domega}), we note that $\omega$ can not be specified uniquely  and to ensure that the integration \eq{MPTS} gives the genuine mass $m$, one can employ the method of Killing-Yano tensor \cite{DK3} (details see also \cite{Killing-yano1,Killing-yano2}) to obtain
\begin{align}\label{ww}
{\bm \omega}=&-\frac n3\,dr\wedge(dt+2n\cos\theta\,d\phi)\n\\
&-\frac r3\sin\theta(r^2+n^2)\,d\theta\wedge d\phi.
\end{align}
From integration (\ref{omegaintegral}), we also notice that $\W_\pm^{r=\inf}$ and $\W_\pm^{r=r_h}$ are determined freely by a constant. Likewise, here we also choose $\W_\pm^{r=\inf}$ only containing the divergent term. By integrating on \eq{ww}, we have
\ba\begin{aligned}
\W^{r=\inf}_\pm=\mp\lim_{r\to\inf}\frac{4\pi n^2}{3}r\,, \quad \W^{r=r_h}_\pm=\mp\frac{4\pi n^2}{3}r_h\,.\\
\end{aligned}\ea
Using above results, we can further obtain the thermodynamic volume
\ba\begin{aligned}
V&=\frac{4\pi r_h^3}{3}\lf(1+\frac{3n^2}{r_h^2}\rt), \\
\end{aligned}\ea
which is in accord with the uncharged case \cite{DK3}. Next, Eq. (\ref{kaba}) gives electric and magnetic potential by
\ba\begin{aligned}
\F_e=h_0-h(r)\,, \ \ \F_m=p(r)-p_0\,.
\end{aligned}\ea
First, we consider the gauge \eq{gauge1}. Under this gauge, we have
\begin{align}
\Phi_e^{(1)}=\frac{2gn^2-er}{r^2+n^2}-\frac{2gn^2-er_h}{r_h^2+n^2},\quad \Phi_m^{(1)}=\frac{n(2gr+e)}{r^2+n^2}. \label{phi1}
\end{align}
Hence, the electric and magnetic potentials of black hole can be written as
\begin{align}
\phi_e=\frac{er_h-2gn^2}{r^2_h+n^2}\,,\ \ \ \phi_m=\frac{n(2gr_h+e)}{r^2_h+n^2}.
\end{align}

After determining the electric charge $Q_e$ and magnetic counterpart $Q^{(+)}_m$, we begin to tackle the Misner terms and obtain that
\begin{align}
\int_{T_+}\star d{\bm k}&=-\frac{8\pi n^2r}{l^2}\bigg|_{r\to\infty}+\frac{8\pi n^2r_h}{l^2}-\frac{2\pi^2(e^2+4g^2n^2)}{n}\n\\
+&\frac{4\pi\left[2mn^2+(e^2+4g^2n^2-2n^2)r_h\right]}{r_h^2+n^2}-\frac{32\pi n^4r_h}{l^2(r_h^2+n^2)}\n\\
+&\frac{4\pi(e^2+4g^2n^2)\arctan\left(\frac{r_h}{n}\right)}{n}.
\end{align}
From this result, we find that $2\Lambda\Omega_+^{r=\infty}=(8\pi n^2r/l^2)\big|_{r\to\infty}$, exactly cancelling out the divergence term. And then, we perform the integrals on field strength tensors over the Misner string under the choice of gauge (\ref{gauge1}) and obtain the following results:
\begin{align}
\int_{T_+}2\Phi^{(1)}_e{\bm G}&=\frac{\pi^2(e^2+4g^2n^2)}{n}-\frac{2\pi r_h(e^2-4g^2n^2)(r_h^2-n^2)}{(r_h^2+n^2)^2} \n\\
+&\frac{16\pi egn^2r^2_h}{(r_h^2+n^2)^2}-\frac{2\pi(e^2+4g^2n^2)\arctan\left(\frac{r_h}{n}\right)}{n} \n\\
-&\frac{8n^2\pi\left[er_h+g(r_h^2-n^2)\right](e+2gr_h)}{(r^2_h+n^2)^2},\n\\
\int_{T_+}2\Phi^{(1)}_m{\bm F}&=-\frac{\pi^2(e^2+4g^2n^2)}{n}+\frac{8\pi gn^2(e+gr_h)(r^2_h-n^2)}{(r_h^2+n^2)^2} \n\\
+&\frac{2\pi e^2r_h(r^2_h+3n^2)}{(r_h^2+n^2)^2}+\frac{2\pi(e^2+4g^2n^2)\arctan\left(\frac{r_h}{n}\right)}{n}. \n
\end{align}
Hence according to Eq. \eq{gmisner}, above calculations yield
\begin{align}
16\pi \y N_+^{(1)}=&\int_{T_+}\left(\star d{\bm k}+\frac{4\Lambda}{n-2}\Omega_+^{(r=\infty)}+2\Phi^{(1)}_e{\bm G}-2\Phi^{(1)}_m{\bm F}\right) \n\\
=&-\frac{4n^2\pi}{r_h}\left[1+\frac{3(n^2-r^2_h)}{l^2}+\frac{(r_h^2-n^2)(e^2+4egr_h)}{(r^2_h+n^2)^2}\right. \n\\
&\qquad\qquad\quad \left.-\frac{4n^2g^2(3r_h^2+n^2)}{(r_h^2+n^2)^2}\right].
\end{align}
By considering $N_+^{(1)}=N_-^{(1)}$ and using $\psi$ in Eq. (\ref{psi}), the Misner charge can be ultimately obtained as
\begin{align}
N^{(1)}=&-\frac{4n^3\pi}{r_h}\left[1+\frac{3(n^2-r^2_h)}{l^2}+\frac{(r_h^2-n^2)(e^2+4egr_h)}{(r^2_h+n^2)^2}\right. \n\\
&\qquad\qquad\quad \left.-\frac{4n^2g^2(3r_h^2+n^2)}{(r_h^2+n^2)^2}\right]\,. \label{N1}
\end{align}
It is now easy to verify that with these quantities the first law is satisfied,
\begin{align}
\delta M=T\,\delta S+\phi_e\,\delta Q_e+\phi_m\,\delta Q^{(+)}_m+\psi\,\delta N^{(1)}+V\,\delta P.
\end{align}

Analogously, under the gauge \eq{gauge2}, the electric and magnetic potentials can be written as
\begin{align}
\Phi^{(2)}_e(r)=\frac{2gn^2-er}{r^2+n^2}, \quad \Phi^{(2)}_m(r)=\frac{n(2gr+e)}{r^2+n^2}-\frac{n(2gr_h+e)}{r_h^2+n^2}. \label{phi2}
\end{align}
Through similar calculations, we obtain that
\begin{align}
\int_{T_+}2\Phi^{(2)}_e{\bm G}&=\frac{\pi^2(e^2+4g^2n^2)}{n}+\frac{2\pi(4egn^2-e^2r_h)(r_h^2-n^2)}{(r_h^2+n^2)^2} \n\\
-&\frac{8\pi g^2n^2r_h(r^2_h+3n^2)}{(r_h^2+n^2)^2}-\frac{2\pi(e^2+4g^2n^2)\arctan\left(\frac{r_h}{n}\right)}{n}, \n\\
\int_{T_+}2\Phi^{(2)}_m{\bm F}&=-\frac{\pi^2(e^2+4g^2n^2)}{n}+\frac{2\pi r_h(4g^2n^2-e^2)(r^2_h-n^2)}{(r_h^2+n^2)^2} \n\\
+&\frac{16\pi egn^2r^2_+}{(r_h^2+n^2)^2}+\frac{2\pi(e^2+4g^2n^2)\arctan\left(\frac{r_h}{n}\right)}{n} \n\\
+&\frac{4\pi[4n^2gr_h-e(r^2_h-n^2)](-2gn^2+er_h)}{(r_h^2+n^2)^2}. \n
\end{align}
Again, according to the Komar integral of Misner charge \eq{gmisner}, we can obtain that
\begin{align}
N^{(2)}=&-\frac{4n^3\pi}{r_h}\left[1+\frac{3(n^2-r^2_h)}{l^2}+\frac{(r_h^2-n^2)(4n^2g^2-4egr_h)}{(r^2_h+n^2)^2}\right. \n\\
&\qquad\qquad\quad \left.-\frac{e^2(3r_h^2+n^2)}{(r_h^2+n^2)^2}\right]\,. \label{N2}
\end{align}
And the corresponding first law is also satisfied:
\begin{align}
\delta M&=T\,\delta S+\phi_e\,\delta Q^{(+)}_e+\phi_m\,\delta Q_m+\psi\,\delta N^{(2)}+V\,\delta P.
\end{align}
The above two procedures are performed based on two different choices of gauge from this geometric perspective. One case is that the electric potential on the horizon and magnetic potential at infinity vanish so that the asymptotic electric charge and horizon magnetic charge are considered as variables of thermodynamics, and the other is just a contrary gauge to regard the asymptotic magnetic charge and horizon electric charge as the independent thermodynamic quantities. Under these two gauges, different Misner charges $N^{(1)}$ and $N^{(2)}$ are obtained by Komar integrations.

\section{general thermodynamic laws with mixed infinity/horizon charges under other gauge choices}\label{5}
The analyses in Sec. \ref{3} and \ref{4} naturally raise an interesting question: could we formulate a more general full cohomogeneity first law and Smarr relation, in which case the electric charge is the mixture of corresponding charges at asymptotic infinity and on the horizon, and so is magnetic charge. We find that it is feasible for general Smarr relation \eq{dec}, since we can always make proper decompositions for the $\Phi_e$ and $\Phi_m$ in arbitrary gauges. However, the consistent first laws will only hold under certain conditions. Next, we first construct general Smarr relation in forms of mixed infinity/horizon charges, and it is not hard to see that the electric potential in any gauge can be decomposed as follows:
\ba\begin{aligned}
\F_e&=\F^{(1)}_e-\a \f_e\\
&=\F^{(1)}_e+\a \lf(\F^{(2)}_e-\F_e^{(1)}\rt)\\
&=(1-\a)\F^{(1)}_e+\a \F^{(2)}_e\,,
\end{aligned}\ea
Similarly for the magnetic potential, we can also choose the gauge as
\ba
\F_m=(1-\b)\F_m^{(1)}+\b \F^{(2)}_m\,.
\ea
Here $\alpha$ and $\beta$ could be either constants or functions of independent variables $\{e,g,r_h,n,l\}$. Thus, the decomposition \eq{dec} can be written as
\begin{align}\label{sm2}
M=&2\lf(TS-PV+\y N^\F\rt)+\phi_e\widetilde{Q}_e+\phi_m\widetilde{Q}_m \,,
\end{align}
where the mixed electric and magnetic charge are defined by
\begin{align}
\widetilde{Q}_e&=\alpha Q^{(+)}_e+(1-\alpha)Q_e, \n\\
\widetilde{Q}_m&=(1-\beta)Q^{(+)}_m+\beta Q_m.
\end{align}
In this case, we can also expressed the novel Misner charge as the mixture of $N^{(1)}$ and $N^{(2)}$.
\begin{align}
N^{\F}=\gamma N^{(1)}+(1-\gamma)N^{(2)}\,,
\end{align}
where $\g$ is a constant which should be determined by demanding the Smarr relation \eq{sm2}. To satisfy this relation, we can deduce that
\begin{align}
&\int_{T_+}\left[\alpha\,\Phi^{(2)}_e{\bm G}+(1-\alpha)\,\Phi^{(1)}_e{\bm G}\right]-\left[(1-\beta)\,\Phi^{(1)}_m{\bm F}+\beta\,\Phi^{(2)}_m{\bm F}\right]\n\\
&=\gamma\int_{T_+}\left(\Phi^{(1)}_e{\bm G}-\Phi^{(1)}_m{\bm F}\right)+(1-\gamma)\int_{T_+}\left(\Phi^{(2)}_e{\bm G}-\Phi^{(2)}_m{\bm F}\right). \n
\end{align}
Further calculation shows that
\begin{align}
&(\alpha+\gamma-1)\int_{T_+}\left(\Phi^{(2)}_e-\Phi^{(1)}_e\right){\bm G} \n\\
=&(\beta+\gamma-1)\int_{T_+}\left(\Phi^{(2)}_m-\Phi^{(1)}_m\right){\bm F}. \n
\end{align}
After performing integrals of two sides, we find that
\begin{align}
&\int_{T_+}\left(\Phi^{(2)}_e-\Phi^{(1)}_e\right){\bm G}=-\int_{T_+}\left(\Phi^{(2)}_m-\Phi^{(1)}_m\right){\bm F} \n\\
=&-\frac{4\pi n^2(2gn^2-er_h)(2gr_h+e)}{(r^2_h+n^2)^2}.
\end{align}
Hence, we obtain that
\begin{align}
\gamma=1-\frac{\alpha+\beta}{2}.
\end{align}
At present, we have accomplished the formulation of general Smarr relation after determining the value of $\gamma$. However, these variables have to satisfy the first law of thermodynamics, namely
\begin{align}
\delta M=T\,\delta S+\phi_e\,\delta\widetilde{Q}_e+\phi_m\,\delta\widetilde{Q}_m+\psi\,\delta\widetilde{N}+V\,\delta P. \label{deltaM}
\end{align}
We can reexpress the first law in terms of independent physical parameters $\{r_h,e,g,l,n\}$. For example, the left hand side becomes
\begin{align}
\delta M=\frac{\p M}{\p r_h}\,\delta r_h+\frac{\p M}{\p e}\,\delta e+\frac{\p M}{\p g}\,\delta g+\frac{\p M}{\p n}\,\delta n+\frac{\p M}{\p l}\,\delta l.\n
\end{align}
Analogous manipulations are made for the right hand side and therefore, Eq. (\ref{deltaM}) is equivalent to the following equations:
\begin{align}
&\frac{n^2}{(r_h^2+n^2)^2}\left\{\left[2eg(n^2-r_h^2)+(4g^2n^2-e^2)r_h\right]\frac{\p}{\p r_h}(\alpha+\beta)\right.\n\\
&\left.\qquad\qquad\quad+(e^2+4g^2n^2)(\alpha-\beta)\right\}=0, \n\\
&\frac{n^2}{(r_h^2+n^2)^2}\left\{\left[(4g^2n^2-e^2)r_h+2eg(n^2-r_h^2)\right]\frac{\p}{\p e}(\alpha+\beta)\right.\n\\
&\left.\qquad\qquad\quad+2g(r_h^2+n^2)(\alpha-\beta)\right\}=0,\n\\
&\frac{n^2}{(r_h^2+n^2)^2}\left\{\left[(4g^2n^2-e^2)r_h+2eg(n^2-r_h^2)\right]\frac{\p}{\p g}(\alpha+\beta)\right.\n\\
&\left.\qquad\qquad\quad-2e(r_h^2+n^2)(\alpha-\beta)\right\}=0, \n\\
&\frac{n(e+2gr_h)}{(r_h^2+n^2)^2}\left[n(2gn^2-er_h)\frac{\p}{\p n}(\alpha+\beta)\right. \n\\
&\left.\qquad\qquad\quad-(er_h+2gn^2)(\alpha-\beta)\right]=0.\n\\
&\frac{n^2}{(r_h^2+n^2)^2}(2gn^2-er_h)(e+2gr_h)\frac{\p}{\p l}(\alpha+\beta)=0. \n
\end{align}
Thus, when $\alpha$ and $\beta$ satisfy the above conditions, we have completed the construction of the first law of thermodynamics. The above five expressions can be further reduced. For simplicity, we can set $x=\alpha+\beta$ and $y=\alpha-\beta$. Then,
\begin{align}\label{75}
\frac{\p x}{\p r_h}&=\frac{(e^2+4g^2n^2)y}{(-2gn^2+er_h)(2gr_h+e)}, \quad \frac{\p x}{\p l}=0. \n\\
\frac{\p x}{\p e}&=\frac{2g(r_h^2+n^2)y}{(-2gn^2+er_h)(2gr_h+e)}, \n\\
\frac{\p x}{\p g}&=\frac{2e(r_h^2+n^2)y}{(2gn^2-er_h)(2gr_h+e)}, \quad \frac{\p x}{\p n}=\frac{(2gn^2+er_h)y}{n(2gn^2-er_h)}. \n
\end{align}
Now, based on these conditions, we can make some discussions. when $\alpha$ is equal to $\beta$, we note that $\alpha$ and $\beta$ are just constants. Specially, we find that if $\alpha=\beta=0$, it demonstrates that the asymptotic electric charge and horizon magnetic counterpart are considered as thermodynamics quantities and general Smarr relation (\ref{sm2}) will be reduced to Eq. (\ref{MT1}), and similarly if $\alpha=\beta=1$, Eq. (\ref{MT2}) are recovered. At the same time, we find that $\alpha=1$, $\beta=0$ or $\alpha=0$, $\beta=1$ can not solve the above equations. Therefore, although the Smarr relations with the electric and magnetic charges at infinity or on the horizon being the thermodynamic quantities can be formulated as follows,
\begin{align}
M&=2(TS-pV+\psi N')+\phi_eQ^{(+)}_e+\phi_mQ^{(+)}_m, \n\\
M&=2(TS-pV+\psi N')+\phi_eQ_e+\phi_mQ_m,
\end{align}
where $N'=(N^{(1)}+N^{(2)})/2$, the corresponding first laws can not be established consistently.

However, for general cases where $\alpha$ and $\beta$ are functions of physical parameters $\{e, g, r_h, n\}$, there exist many solutions for above equations, which correspond to different forms of thermodynamics under arbitrary choices of gauge, and here we only find some special ones. For instance, when $\beta$ is equal to some constant, we can obtain that
\begin{align}
\frac{\p y}{\p r_h}&=\frac{(e^2+4g^2n^2)y}{(-2gn^2+er_h)(2gr_h+e)}, \frac{\p y}{\p g}=\frac{2e(r_h^2+n^2)y}{(2gn^2-er_h)(2gr_h+e)}, \n\\
\frac{\p y}{\p e}&=\frac{2g(r_h^2+n^2)y}{(-2gn^2+er_h)(2gr_h+e)}, \frac{\p y}{\p n}=\frac{(2gn^2+er_h)y}{n(2gn^2-er_h)}.\n
\end{align}
It is not difficult to find that the final result is
\begin{align}
\alpha-\beta=C_1\frac{er_h-2gn^2}{n(e+2gr_h)}, \quad \beta=C_2.
\end{align}
Analogously, if we take $\alpha$ to be a constant, then
\begin{align}
\alpha-\beta=C_3\frac{n(e+2gr_h)}{er_h-2gn^2}, \quad \alpha=C_4.
\end{align}
Here, $C_1$, $C_2$, $C_3$ and $C_4$ are all arbitrary constants. Obviously, for charged Taub-NUT-AdS spacetime containing both electric and magnetic charges, the first laws of thermodynamics and Smarr relations can be constructed in myriad forms consistently, rather than only two kinds of them as mentioned in the previous sections.

\section{Conclusions}\label{6}
As shown by some recent studies \cite{Clement1,Clement2,Clement3}, Taub-NUT solution in Lorentzian setup does not lead to serious problems as imagined previously, and when establishing the Smarr relation and first law of thermodynamics, the core ingredient is to introduce a pair of novel conjugate thermodynamic quantities, namely the Misner potential and charge $\psi$--$N$ \cite{DK1,DK3} to ensure that the NUT parameter $n$ varies independently. The thermodynamics of the uncharged Lorentzian Taub-NUT-AdS maintaining the Misner strings present has been formulated, and from the Komar integral notions, it turns out that $\psi$ is actually the surface gravity of the Misner strings and $N$ is interpreted as the integration over it \cite{DK3}. When extended to charged case, due to this spacetime's peculiar property as described in the previous sections, two different types of thermodynamics have been constructed by calculating Euclidean action for free energy to determine the Misner charge and other thermodynamic quantities \cite{DK2}.

Inspired by the uncharged case \cite{DK3} and solution with rotation \cite{DK4}, in this paper we have reformulated the thermodynamic laws of the charged NUT black hole with the Misner strings present in the Einstein-Maxwell gravity from the Komar integral perspective, and shows that the Misner charges in this case are also corresponding to Komar integrals over the Misner strings. Utilizing the language of differential forms, we give the universal Smarr relation (\ref{dec}) directly. As a result of distinctive property of the electric and magnetic charges illustrated in (\ref{qeqm}), the gauge freedom of the electric and magnetic potentials is non-trivial, which demonstrates that there is bound to exist a myriad of Smarr relations and meanwhile, we give the necessary conditions to satisfy the first laws consistently.  Based on this philosophy, we argue that the two types of thermodynamics discussed in \cite{DK2} are just the special cases of Eq. (\ref{dec}) under two specified gauge choices. And then, we extend this notion to more general thermodynamics when choosing other arbitrary gauges and thus obtain two novel non-trivial versions of thermodynamic laws. It is worth noting that the strategy of constructing thermodynamics of Nutty dyonic black hole in Ref. \cite{DK2} is to evaluate Euclidean action to give the expression of free energy and then identify different thermodynamic conjugate pairs, but the determination of electric and magnetic terms is fully out of physical consideration and this technique is unconducive to generalizing to more general thermodynamics with the mixed horizon/infinity charges. Instead, the geometric perspective gives corresponding Komar integrals of different conjugate pairs directly and it is obvious that distinct forms of thermodynamics will be obtained only by adjustment of the electric and magnetic potentials under certain conditions. {The above results are completely different from those of the normal black holes, where the first laws of thermodynamics are not dependent on the gauge choices. However, this can be understood by the non-trivial topology of the NUT black holes. For a normal black hole, there are only two boundaries of the hypersurface connecting the horizon and asymptotic infinity, and the evaluation of the electric and magnetic charges on the horizon are the same as these at infinity. Then, the decomposition of the black hole mass will be independent on the gauge choice. While for the NUT black hole, by virtue of the existence of the Misner string, the evaluation of the electric and magnetic charges will depend on the sphere radius, which will make the decomposition rely on the gauge choice.

}

In summary, geometric perspective provides us a straightforward means to establish universal thermodynamic laws for Nutty dyonic black hole. However, many problems are still left to be further investigation. For instance, more intriguing forms of mixed charges need to be constructed, the physical meaning of Misner charge is still obscure and for other dyonic black hole, whether we can establish different types of thermodynamics from the geometric notion. These deserve our deep explorations.
\\

\section*{acknowledgements}
This research was supported by National Natural Science Foundation of China (NSFC) with Grants No. 11375026, No. 11675015, and No. 11775022.


\begin{thebibliography}{99}
\bibitem{Taub}
  A.~H.~Taub, ``Empty space-times admitting a three parameter group of motions,''  Annals Math.\  {\bf 53}, 472 (1951).
\bibitem{Newman}
E. Newman, L. Tamburino, and T. Unti, ``Empty-space generalization of the Schwarzschild metric,'' J.\ Math.\ Phys.\  {\bf 4}, 915 (1963).
\bibitem{CWM}
  C.~W.~Misner, ``The Flatter regions of Newman, Unti and Tamburino's generalized Schwarzschild space,'' J.\ Math.\ Phys.\  {\bf 4}, 924 (1963).

\bibitem{Hawking}
S. W. Hawking and G. F. R. Ellis, ``The large scale structure of space-time, vol. 1.'' Cambridge university press, 1973.

\bibitem{Hajicek}
P. Hajicek, Causality in non-Hausdorff space-times, $Communications$ $in$ $Mathematical$ $Physics$ {\bf 21} (1971) 75.

\bibitem{Stelea}
R. B. Mann and C. Stelea, ``On the thermodynamics of NUT charged spaces,'' Phys.\ Rev.\ D {\bf 72} 084032 (2005).

\bibitem{Johnson1}
C. V. Johnson, ``Thermodynamic Volumes for AdS-Taub-NUT and AdS-Taub-Bolt,'' Class. Quant. Grav. {\bf 31} 235003 (2014).

\bibitem{Johnson2}
C. V. Johnson, ``The Extended Thermodynamic Phase Structure of Taub-NUT and Taub-Bolt,'' Class. Quant. Grav. {\bf 31} 225005 (2014).

\bibitem{Clement1}
G. Cl\'{e}ment, D. Gal'tsov and M. Guenouche, ``Rehabilitating space-times with NUTs,'' Phys. Lett. B {\bf750} 591 (2015).

\bibitem{Clement2}
G. Cl\'{e}ment, D. Gal'tsov and M. Guenouche, ``NUT wormholes,'' Phys. Rev. D {\bf93} 024048 (2016).

\bibitem{Clement3}
G. Cl\'{e}ment and M. Guenouche, ``Motion of charged particles in a NUTty Einstein-Maxwell spacetime and causality violation,'' Gen. Rel. Grav. {\bf 50} 60 (2018).

\bibitem{DK1}
R. A. Hennigar, D. Kubiznak, and R. B. Mann, ``Thermodynamics of Lorentzian Taub-NUT spacetimes,'' arXiv:1903. 08668.

\bibitem{Brown1}
 J. D. Brown and C. Teitelboim, ``Dynamical Neutralization of the Cosmological Constant,'' Phys. Lett. B {\bf195}, 177 (1987).

\bibitem{Brown2}
J. D. Brown and C. Teitelboim, ``Neutralization of the Cosmological Constant by Membrane Creation,'' Nucl. Phys. B {\bf297}, 787 (1988).

\bibitem{Emparan}
R. Emparan, C. V. Johnson, and R. C. Myers, ``Surface terms as counterterms in AdS/CFT correspondence,'' Phys. Rev. D {\bf 60} 104001 (1999).

\bibitem{DK3}
A. B. Bordo, F. Gray, R. A. Hennigar and D. Kubiznak, ``Misner Gravitational Charges and Variable String Strengths,'' arXiv:1905.03785.

\bibitem{Brill}
D. R. Brill, ``Electromagnetic fields in a homogeneous, nonisotropic universe,'' Phys. Rev. {\bf 133} B845 (1964).

\bibitem{Demianski}
J. F. Plebanski and M. Demianski, ``Rotating, charged, and uniformly accelerating mass in general relativity,'' Annals Phys.\  {\bf 98}, 98 (1976).

\bibitem{NA}
N.~Alonso-Alberca, P.~Meessen and T.~Ortin, ``Supersymmetry of topological Kerr-Newman-Taub-NUT-AdS space-times,'' Class.\ Quant.\ Grav.\  {\bf 17}, 2783 (2000).

\bibitem{DK2}
A. B. Bordo, F. Gray, and D. Kubiznak, ``Thermodynamics and Phase Transitions of NUTty Dyons,'' arXiv:1904.00030.

\bibitem{AS}
A. Ashtekar and S. Das, ``Asymptotically Anti-de Sitter space-times: Conserved quantities,'' Class. Quant. Grav. {\bf 17}  L17 (2000).

\bibitem{DK4}
A. B. Bordo, F. Gray, R. A. Hennigar, and D. Kubiznak, ``The First Law for Rotating NUTs,'' arXiv:1905.06350.

\bibitem{Kastor}
D. Kastor, S. Ray, and J. Traschen, ``Enthalpy and the Mechanics of AdS Black Holes'', Class. Quant. Grav. {\bf 26} 195011 (2009).

\bibitem{Killing-yano1}
D. Kubiznak and P. Krtous,``On conformal Killing-Yano tensors for Plebanski-Demianski family of solutions'', Phys. Rev. D {\bf76} 084036 (2007).

\bibitem{Killing-yano2}
V. Frolov, P. Krtous, and D. Kubiznak, Black holes, hidden symmetries, and complete integrability, Living Rev. Rel. {\bf 20} 16 (2017).



\end{thebibliography}
\end{document}